%% file: main.tex
\documentclass[letterpaper]{article} % DO NOT CHANGE THIS
\usepackage{aaai23}  % DO NOT CHANGE THIS
\usepackage{times}  % DO NOT CHANGE THIS
\usepackage{helvet}  % DO NOT CHANGE THIS
\usepackage{courier}  % DO NOT CHANGE THIS
\usepackage[hyphens]{url}  % DO NOT CHANGE THIS
\usepackage{graphicx} % DO NOT CHANGE THIS
\usepackage{natbib}  % DO NOT CHANGE THIS AND DO NOT ADD ANY OPTIONS TO IT
\usepackage{caption} % DO NOT CHANGE THIS AND DO NOT ADD ANY OPTIONS TO IT
\usepackage{graphicx} % for \resize
\usepackage{bm} % for bm
\usepackage{amsmath} % for \aligned \underbrace \mathbf
\usepackage{amsfonts} % for \mathbb
\usepackage{algorithm} % for algorithm
\usepackage{algpseudocode} % for algorithm
\usepackage{tablefootnote} % for \tablefootnote
\usepackage{multirow} % for \multirow
\usepackage{enumitem} % for itemsep
\usepackage{booktabs}  % for \toprule 

\usepackage{newfloat}
\usepackage{listings}
\DeclareCaptionStyle{ruled}{labelfont=normalfont,labelsep=colon,strut=off} % DO NOT CHANGE THIS
\floatstyle{ruled}
\newfloat{listing}{tb}{lst}{}
\floatname{listing}{Listing}
\graphicspath{{./images/}}
\title{Molformer: Motif-based Transformer on 3D Heterogeneous Molecular Graphs}
\author{
    Fang Wu,\textsuperscript{\rm 1}\textsuperscript{\rm 3}
    Dragomir Radev, \textsuperscript{\rm 2}
    Stan Z. Li \textsuperscript{{\rm 1}}\thanks{The corresponding author.}
}
\affiliations{
    \textsuperscript{\rm 1} School of Engineering, Westlake University \\
    \textsuperscript{\rm 2} Department of Computer Science, Yale University \\
    \textsuperscript{\rm 3} Institute of AI Industry Research, Tsinghua University \\ 
    fw2359@columbia.edu, dragomir.radev@yale.edu, stan.zq.li@westlake.edu.cn
}

\usepackage{bibentry}
\begin{document}
\maketitle
\begin{abstract}
Procuring expressive molecular representations underpins AI-driven molecule design and scientific discovery. The research mainly focuses on atom-level homogeneous molecular graphs, ignoring the rich information in subgraphs or motifs. However, it has been widely accepted that substructures play a dominant role in identifying and determining molecular properties.
To address such issues, we formulate heterogeneous molecular graphs (HMGs), and introduce a novel architecture to exploit both molecular motifs and 3D geometry. Precisely, we extract functional groups as motifs for small molecules and employ reinforcement learning to adaptively select quaternary amino acids as motif candidates for proteins. Then HMGs are constructed with both atom-level and motif-level nodes. To better accommodate those HMGs, we introduce a variant of the Transformer named Molformer, which adopts a heterogeneous self-attention layer to distinguish the interactions between multi-level nodes. Besides, it is also coupled with a multi-scale mechanism to capture fine-grained local patterns with increasing contextual scales. An attentive farthest point sampling algorithm is also proposed to obtain the molecular representations. We validate Molformer across a broad range of domains, including quantum chemistry, physiology, and biophysics. Extensive experiments show that Molformer outperforms or achieves the comparable performance of several state-of-the-art baselines. Our work provides a promising way to utilize informative motifs from the perspective of multi-level graph construction. The code is available at~\url{https://github.com/smiles724/Molformer}.
\end{abstract}
\input{1_intro}
\input{2_preliminary}
\input{3_heterogeneous}
\input{4_molformer}
\input{5_experiment}
\input{6_related_work}
\input{7_conclusion}

\newpage
\bibliography{aaai23}

\newpage
\input{8_appendix}

\end{document}

%% file: 1_intro.tex
\section{Introduction}
The past decade has witnessed the extraordinary success of deep learning (DL) in many scientific areas. Inspired by these achievements, researchers have shown increasing interest in exploiting DL for drug discovery and material design with the hope of rapidly identifying desirable molecules. A key aspect of fast screening is how to represent molecules effectively, where graphs are a natural choice to preserve their internal structures. As a consequence, a number of \emph{graph neural networks} (GNNs)~\citep{gilmer2017neural,ishida2021graph} have been invented and applied to molecular representation learning with noticeable performance.   % on property prediction. 

However, most existing GNNs only take atom-level information in homogeneous molecular graphs as input, which fails to adequately exploit rich semantic information in motifs. Remarkably, motifs are significant subgraph patterns that frequently occur, and can be leveraged to uncover molecular properties. For instance, a carboxyl group (\emph{COOH}) acts as a hydrogen-bond acceptor, which contributes to better stability and higher boiling points. Besides, similar to the role of N-gram in natural language, molecular motifs can promote the segmentation of atomic semantic meanings. While some regard motifs as additional features of atoms~\citep{maziarka2020molecule,maziarka2021relative}, 
these methods increase the difficulty to separate the semantic meanings of motifs from atoms explicitly and hinder models from viewing motifs from an integral perspective. Others~\citep{huang2020caster} take motifs as the only input, but ignore the influence of single atoms and infrequent substructures. 

To overcome these problems, we formulate a novel heterogeneous molecular graph (HMG) that is comprised of both atom-level and motif-level nodes as the model input. It provides a clean interface to incorporate nodes of different levels and prevents the error propagation caused by incorrect semantic segmentation of atoms. As for the determination of motifs, we adopt different strategies for different types of molecules. On the one hand, for small molecules, the motif lexicon is defined by functional groups based on chemical domain knowledge. On the other hand, for proteins that are constituted of sequential amino acids, a reinforcement learning (RL) motif mining technique is introduced to discover the most meaningful amino acid subsequences for downstream tasks. 

In order to better align with HMGs, we present Molformer, an equivariant geometric model based on Transformer~\citep{vaswani2017attention}. Molformer differs from preceding Transformer-based models in two major aspects. First, it utilizes heterogeneous self-attention (HSA) to distinguish the interactions between nodes of different levels and incorporates them into the self-attention computation. Second, an Attentive Farthest Point Sampling (AFPS) algorithm is introduced to aggregate node features and obtain a comprehensive representation of the entire molecule.

To summarize, our contributions are as follows: 
\begin{itemize} % [noitemsep]
    \item To the best of our knowledge, we are the foremost to incorporate motifs and construct 3D heterogeneous molecular graphs for molecular representation learning. 
    \item We propose a novel Transformer architecture to perform on these heterogeneous molecular graphs. It has a modified self-attention to take into account the interactions between multi-level nodes, and an AFPS algorithm to integrate molecular representations. 
    \item We empirically outperform or achieve competitive results compared to state-of-the-art baselines on several benchmarks of small molecules and proteins.
\end{itemize}

%% file: 2_preliminary.tex
\section{Preliminaries}
\paragraph{Problem Definition.}
Suppose a molecule $\boldsymbol{S}=(\boldsymbol{P}, \boldsymbol{H})$ has $N$ atoms, where $\boldsymbol{P} = \{\boldsymbol{p}_i\}_{i=1}^N\in \mathbb{R}^{N \times 3}$ describes 3D coordinates associated to each atom and $\boldsymbol{H} =\{\boldsymbol{h}_i\}_{i=1}^N\in \mathbb{R}^{N \times h}$ contains a set of $h$-dimension roto-translationally invariant features (e.g. atom types and weights).
$\boldsymbol{h}_i$ can be converted to a dense vector $\boldsymbol{x}_i \in \mathbb{R}^{\psi_\textrm{embed}}$ via a multi-layer perceptron (MLP).
A representation learning model $f$ acts on $\boldsymbol{S}$, and obtains its representation $\boldsymbol{r}=f(\boldsymbol{S})$. Then $\boldsymbol{r}$ is forwarded to a predictor $g$ and attains the prediction of a biochemical property $\hat{y}=g(\boldsymbol{r})$.

\paragraph{Self-Attention Mechanism.} 
Given input ${\{\boldsymbol{x}_i\}}_{i=1}^N$, the standard dot-product self-attention layer is computed as follows:
\begin{align}
    \boldsymbol{q}_i = f_Q(\boldsymbol{x}_i), \, \boldsymbol{k}_i = f_K(\boldsymbol{x}_i), \, \boldsymbol{v}_i = f_V(\boldsymbol{x}_i) \\ 
    a_{ij} = \boldsymbol{q}_i\boldsymbol{k}_j^T/\sqrt{\psi_\textrm{model}}, \,
    \boldsymbol{z}_i = \sum_{j=1}^N\sigma(a_{ij})\boldsymbol{v}_j
\end{align}
where $\{f_Q, f_K, f_V\}$ are embedding transformations, and $\{\boldsymbol{q}_i, \boldsymbol{k}_i, \boldsymbol{v}_i\}$ are respectively the query, key, and value vectors with the same dimension $\psi_\textrm{model}$. $a_{ij}$ is the attention that the token $i$ pays to the token $j$. $\sigma$ denotes the \textit{Softmax} function and $\boldsymbol{z}_i$ is the output embedding of the token $i$. 
This formula conforms to a non-local network~\citep{wang2018non}, which indicates its poor ability to capture fine-grained patterns in a local context. 

\paragraph{Position Encoding.} 
Self-attention is invariant to the permutation of the input, and position encoding (PE) is the only technique to reveal position information. PE can be based on absolute positions or relative distances. The former uses raw positions and is not robust to spatial transformations. The latter manipulates the attention score by incorporating relative distances: $a_{ij}=\boldsymbol{q}_i\boldsymbol{k}_j^T/\sqrt{\psi_\textrm{model}}+f_{\rm{PE}}(\boldsymbol{p}_i-\boldsymbol{p}_j)$, where $f_{\rm{PE}}$ is a translation-invariant PE function. The rotation invariance can be further accomplished by taking a L2-norm $d_{ij}=||\boldsymbol{p}_i-\boldsymbol{p}_j||_2$ between the $i$-{th} and $j$-{th} atom. 

%% file: 3_heterogeneous.tex
\section{Heterogeneous Molecular Graphs}
Motifs are frequently-occurring substructure patterns and serve as the building blocks of complex molecular structures. They have great expressiveness of the biochemical characteristics of the whole molecules. In this section, we first describe how to extract motifs from small molecules and proteins respectively, and then present how to formulate HMGs.
\begin{figure}[t]
\centering
\includegraphics[scale=0.7]{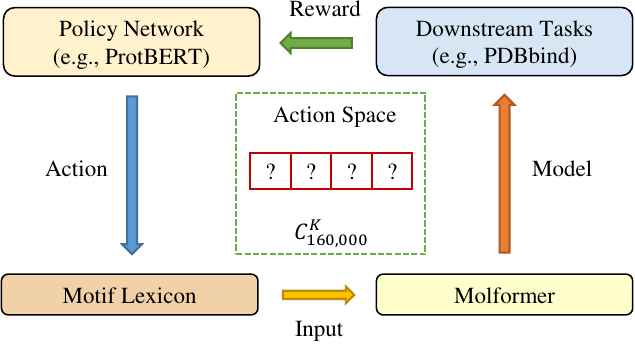}
% \vspace{-1em}
\caption{\textbf{The workflow of RL motif mining method in proteins.} In each iteration, the policy network is responsible for producing a motif lexicon. Molformer's performance on downstream tasks and the diversity of the lexicon are considered as the reward.}
\label{policy_motif}
% \vspace{-1em}
\end{figure}

\subsection{Motifs in Small Molecules}
In the chemical community, a set of standard criteria have been developed to recognize motifs with essential functionalities in small molecules.
There, we build motif templates of four categories of functional groups, containing groups that contain only carbon and hydrogen (Hydrocarbons), groups that contain halogen (Haloalkanes), groups that contain oxygen, and groups that contain nitrogen (see Appendix for more discussion). Practically, we rely on RDKit~\citep{landrum2013rdkit} to draw them from SMILES of small molecules.

\subsection{Motifs in Proteins}
In large protein molecules, motifs are local regions of 3D structures or amino acid sequences shared among proteins that influence their functions~\citep{somnath2021multi}. Each motif usually consists of only a few elements, such as the 'helix-turn-helix' motif, and can describe the connectivity between secondary structural elements. The detection of protein motifs has been long studied. Nevertheless, existing tools are either from the context of a protein surface~\citep{somnath2021multi} or are task-independent and computationally expensive~\citep{mackenzie2016tertiary}. On the basis of this peculiarity, we design an RL mining method to  discover task-specific protein motifs heuristically.

Therefore, we consider motifs with four amino acids because they make up the smallest polypeptide and have special functionalities in proteins. For instance, $\beta$-turns are composed of four amino acids and are a non-regular secondary structure that causes a change in the direction of the polypeptide chain. Each amino acid can be of 20 different possibilities, such as Alanine, Isoleucine, and Methionine, so there are $1.6 \times 10^5$ ($=20^4$) potential quaternary motifs. 

Our goal is to find the most effective lexicon $\boldsymbol{\mathcal{V}}^*\in \mathbb{V}$ composed of $K$ quaternary amino acid templates, where $\mathbb{V}$ denotes the space of all $C_{1.6 \times 10^5}^{K}$ potential lexicons. Since we aim to mine the optimal task-specific lexicon, it is practically feasible to only consider the existing quaternions in the downstream datasets instead of all $1.6 \times 10^5$ possible quaternions.

In each iteration of the parameter update, we use a pre-trained ProtBERT~\cite{elnaggar2020prottrans} with an MLP as the policy network $\pi_{\theta}$. Specifically, all possible quaternions are fed into ProtBert to obtain their corresponding representations $\{\boldsymbol{e}_i\}_{i=1}^{1.6 \times 10^5}$, which are subsequently sent to the MLP to acquire their scores $\{s_i\}_{i=1}^{1.6 \times 10^5}$. These scores illustrate each quaternion's significance to benefit the downstream tasks if they are chosen as a part of the vocabulary. Then top-$K$ motifs with the highest scores are selected to comprise $\boldsymbol{\mathcal{V}}\in \mathbb{V}$ in accordance with $\{s_i\}_{i=1}^{1.6 \times 10^5}$, and $\boldsymbol{\mathcal{V}}$ is used as templates to extract motifs and construct HMGs in downstream tasks. After that, a Molformer is trained based on these HMGs. Its validation performance is regarded as the reward $r$ to update parameters ${\theta}$ by means of policy gradients. With adequate iterations, the agent can select the optimal task-specific quaternary motif lexicon $\boldsymbol{\mathcal{V}}^*$.

Remarkably, our motif mining process is a one-step game, since the policy network $\pi_{\theta}$ only generates the vocabulary $\boldsymbol{\mathcal{V}}$ once in each iteration. Thus, the trajectory consists of only one action, and the performance of Molformer based on the chosen lexicon $\boldsymbol{\mathcal{V}}$ composes a part of the total reward. Moreover, we also consider the diversity of motif templates within the lexicon and calculate it as:
\begin{equation}
    d_{\textrm{div}}(\boldsymbol{\mathcal{V}}) = \frac{1}{|\boldsymbol{\mathcal{V}}|}\sum_{\boldsymbol{m}_i\in \boldsymbol{\mathcal{V}}}\sum_{\boldsymbol{m}_j\in \boldsymbol{\mathcal{V}}}d_{\textrm{lev}}(\boldsymbol{m}_i, \boldsymbol{m}_j)
\end{equation}
where $d_{\textrm{lev}}$ is the Levenshtein distance of two quaternary sequences $\boldsymbol{m}_i$ and $\boldsymbol{m}_j$. The final reward, therefore, becomes $R(\boldsymbol{\mathcal{V}}) = r + \gamma d_{\textrm{div}}(\boldsymbol{\mathcal{V}})$, where $\gamma$ is a weight to balance two reward terms, and the policy gradient is computed as the following objective:
\begin{equation}
    \nabla_{\theta} J(\theta)=\mathbb{E}_{\boldsymbol{\mathcal{V}}\in \mathbb{V}}[\nabla_{\theta}\log\pi_{\theta}(\boldsymbol{\mathcal{V}})R(\boldsymbol{\mathcal{V}})]
\end{equation}
\subsection{Formulation of Heterogeneous Molecular Graphs}
Most prior studies~\citep{maziarka2020molecule,rong2020self,maziarka2021relative} simply incorporate motifs into atom features. For instance, they differentiate carbons into aromatic or non-aromatic and deem them as extra features. We argue its ineffectiveness for two reasons. First, the fusion of multi-level features increases the difficulty to summarize the functionality of motifs. Second, it hinders models to see motifs from a unitary perspective. To fill these gaps, we separate apart motifs and atoms, regarding motifs as new nodes to formulate an HMG. This way disentangles motif-level and atom-level representations, thus alleviating the difficulty for models to properly mine the motif-level semantic meanings.

Similar to the relation between phrases and single words in natural language, motifs in molecules carry higher-level semantic meanings than atoms. Therefore, they play an essential part in identifying the functionalities of their atomic constituents. Inspired by the employment of dynamic lattice structures in named entity recognition, we treat each category of motifs as a new type of node and build HMGs as the input of our Molformer. 
To begin with, motifs are extracted according to a motif vocabulary $\boldsymbol{\mathcal{V}}$. We assume $M$ motifs $\left\{\boldsymbol{m}_i\right\}_{i=1}^{M}$ are detected in the molecule $\boldsymbol{S}$.
Consequently, an HMG includes both the motif-level and atom-level nodes as $\{\boldsymbol{x}_1,...,\boldsymbol{x}_N,\boldsymbol{x}_{\boldsymbol{m}_1},...,\boldsymbol{x}_{\boldsymbol{m}_M}\}$, where $\boldsymbol{x}_{\boldsymbol{m}_i}\in \mathbb{R}^{\psi_\textrm{embed}}$ is obtained through a learnable embedding matrix $\boldsymbol{W}^M\in \mathbb{R}^{C'\times \psi_\textrm{embed}}$ and $C'$ denotes the number of motif categories. As for positions of each motif, we adopt a weighted sum of 3D coordinates of its components as $\boldsymbol{p}_{\boldsymbol{m}_i}=\sum_{\boldsymbol{x}_i\in \boldsymbol{m}_i}\left(\frac{w_i}{\sum_{\boldsymbol{x}_i\in \boldsymbol{m}_i} w_i}\right) \cdot \boldsymbol{p}_i$, where $w_i$ are the atomic weights. 
Analogous to word segmentation, our HMGs composed of multi-level nodes avoid error propagation due to inappropriate semantic segmentation while leveraging the atom information for molecular representation learning.

%% file: 4_molformer.tex
\section{Molformer}
In this section, we propose Molformer, which modifies Transformer with several novel components specifically designed for 3D HMGs. First, we build a motif vocabulary and match each molecule with this lexicon to obtain all its contained motifs. Then both atoms and motifs acquire their corresponding embeddings and are forwarded into $L$ feature learning blocks. Each block consists of an HSA, a feed-forward network (FFN), and two-layer normalizations. After that, an AFPS is followed to adaptively produce the molecular representation, which is later fed into a dense predictor to forecast properties in a broad range of downstream tasks. 

\begin{figure*}[ht]
\centering
\includegraphics[scale=0.55]{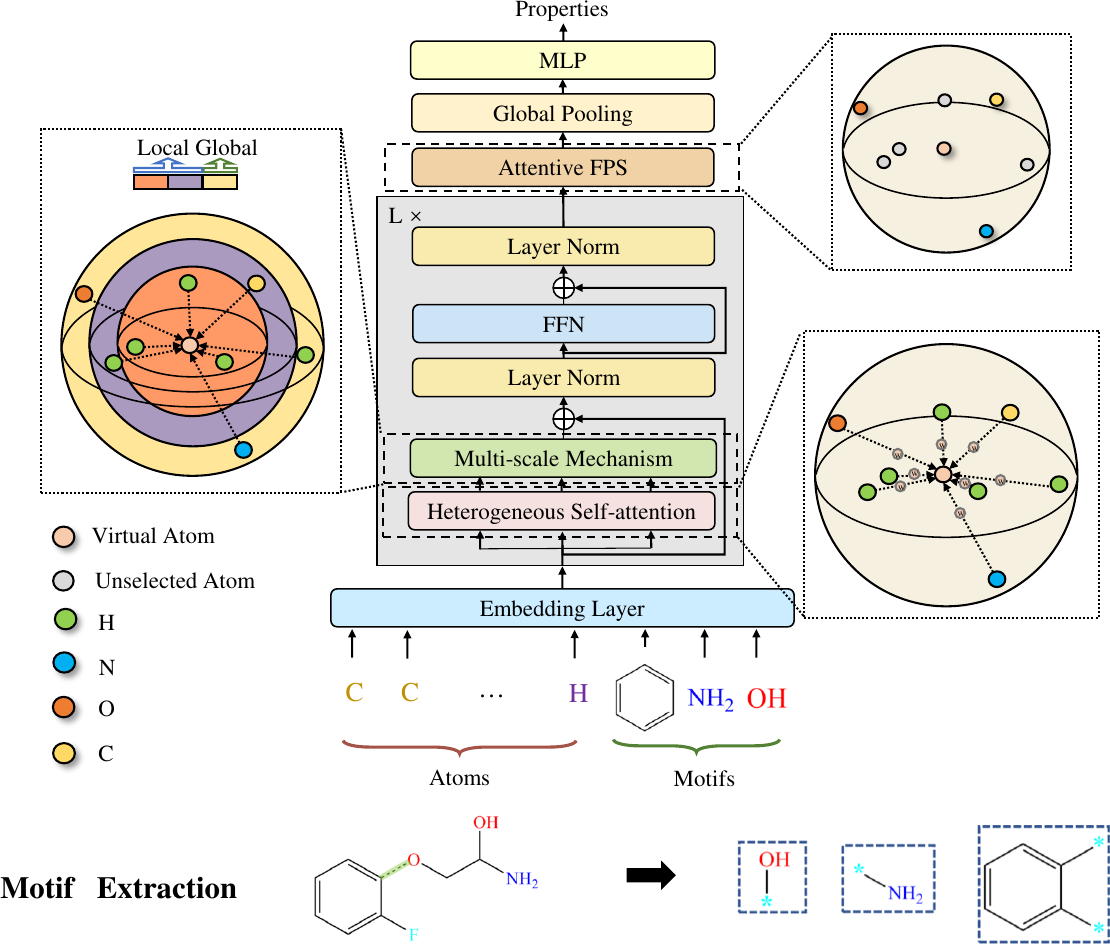}
\caption{\textbf{The architecture of Molformer.} Given a heterogeneous molecular graph with both atom-level and motif-level nodes, stacked feature learning blocks composed of a heterogeneous self-attention module and an FFN compute their updated features. Afterward, an attentive subsampling module integrates the molecular representation for downstream predictions. Local features with different scales are purple and orange; yellow corresponds to global features.}
\label{model}
% \vspace{-1em}
\end{figure*}
\subsection{Heterogeneous Self-attention}
After formulating an HMG with $N$ atom-level nodes and $M$ motif-level nodes, it is important to endow the model with the capacity of separating the interactions between multi-order nodes. To this end, we utilize a function $\phi(i,j): \mathbb{R}^{(N + M)\times (N + M)} \rightarrow \mathbb{Z}$, which identifies the relationship between any two nodes into three sorts: atom-atom, atom-motif, and motif-motif. Then a learnable scalar $b_{\phi(i,j)}: \mathbb{Z} \rightarrow \mathbb{R}$ indexed by $\phi(i,j)$ is introduced so that each node can adaptively attend to all other nodes according to their hierarchical relationship inside our HMGs.

In addition, we consider exploiting 3D molecular geometry (see Figure~\ref{model}). Since robustness to global changes such as 3D translations and rotations is an underlying principle for molecular representation learning, we seek to satisfy roto-translation invariance. There, we borrow ideas from SE(3)-Transformer~\citep{schutt2021equivariant} and AlphaFold2~\citep{jumper2021highly}, and apply a convolutional operation to the pairwise distance matrix $\boldsymbol{D} = [{d}_{i,j}]_{i,j\in[N + M]} \in \mathbb{R}^{(N + M)\times (N + M)}$ as $\hat{\boldsymbol{D}} = \textrm{Conv}_{2d}(\boldsymbol{D})$, where $\textrm{Conv}_{2d}$ denotes a 2D shallow convolutional network with a kernel size of $1\times1$. Consequently, the attention score is computed as follows:
\begin{equation}
\label{equ: cpe}
    \hat{a}_{ij} = \left(\boldsymbol{q}_i\boldsymbol{k}_j^T/\sqrt{\psi_\textrm{model}}\right) \cdot \hat{d}_{ij} + b_{\phi(i,j)},
\end{equation} 
where $\hat{d}_{i,j} \in \hat{\boldsymbol{D}}$ controls the impact of interatomic distance over the attention score, and $b_{\phi(i,j)}$ is shared across all layers. 

Moreover, exploiting local context has proven to be important in sparse 3D space. However, it has been pointed out that self-attention is good at capturing global data patterns but ignores local context~\citep{wu2020lite}.  Based on this fact, we impose a distance-based constraint on self-attention in order to extract multi-scaled patterns from both local and global contexts.
\citet{guo2020multi} propose to use integer-based distances to limit attention to local word neighbors, which cannot be used in molecules. This is because different types of molecules have different densities and molecules of the same type have different spatial regularity, which results in the non-uniformity of interatomic distances. Normally, small molecules have a mean interatomic distance of 1-2 \textup{\AA} (Angstrom, $10^{-10}m$), which is denser than large molecules like proteins with approximately 5 \textup{\AA} on average. To tackle that, we design a multi-scale methodology to robustly capture details. Specifically, we mask nodes beyond a certain distance $\tau_s$ at each scale $s$. The attention calculation is modified as:
\begin{equation} 
\label{equ: msa}
    a_{ij}^{\tau_s} = \hat{a}_{ij} \cdot \mathbf{1}_{\{d_{ij}<\tau_s\}}, \boldsymbol{z}_i^{\tau_s} = \sum_{j=1}^N \sigma\left(a_{ij}^{\tau_s} \right)\boldsymbol{v}_j,
\end{equation}
where $\mathbf{1}_{\{d_{ij}<\tau_s\}}$ is the indicator function. Notably, Equation~\ref{equ: msa} can be complementarily combined with Equation~\ref{equ: cpe}. Then, features extracted from $S$ different scales $\{\tau_s\}_{s=1}^S$, as well as the informative global feature, are concatenated together to form a multi-scale representation, denoted by $\boldsymbol{z}'_i = \boldsymbol{z}_i^{\tau_1} \oplus ... \oplus \boldsymbol{z}_i^{\tau_S} \oplus \boldsymbol{z}_i^{global} \in \mathbb{R}^{(S+1)\psi_\textrm{model}}$. After that, $\boldsymbol{z}'_i$ is forwarded into a FFN to 
obtain $\boldsymbol{z}''_i$ with the original dimension $\psi_\textrm{model}$. 

% With multi-headed self-attention, $\hat{\boldsymbol{D}}$ is expanded in the sense that $\hat{\boldsymbol{D}} \in \mathbb{R}^{ H \times (N + M)\times (N + M)}$, and $\textrm{Conv}_{2d}$ has $H$ output channels. The CPE method induces $\textrm{O}(N+M)$ convolution operations on each atom and can drastically reduce training time when the number of atoms is very large~\citep{wu2021spatial}.

\subsection{Attentive Farthest Point Sampling}
After having the node embeddings $\{\boldsymbol{z}''_i\}_{i=1}^{N+M}$, we study how to obtain the molecular representation $\boldsymbol{r}$. For GNNs, several readout functions such as set2set~\citep{vinyals2015order} and GG-NN~\citep{gilmer2017neural} are invented. For Transformers, one way is via a virtual node. Though~\citet{ying2021transformers} state that it significantly improves the performance of existing models in the leaderboard of Open Graph Benchmark (OGB), this way concentrates more on its close adjacent nodes and less on distant ones, and may lead to inadvertent over-smoothing of information propagation. Besides, it is difficult to locate a virtual node in 3D space and build connections to existing vertices. The other way selects a subset of nodes via a downsampling algorithm named Farthest Point Search (FPS), but it ignores nodes' differences and has the sensitivity to outlier points as well as uncontrollable randomness. To address these issues, we propose a new algorithm named AFPS. It aims to sample vertices by not merely spatial distances, but also their significance in terms of attention scores. 
\begin{algorithm}[t]
\caption{Attentive Farthest Point Sampling}
\label{agl: AFPS_algo}
\begin{algorithmic}
    \State {\bfseries Input:} The attention score matrix $\boldsymbol{A} \in \mathbb{R}^{(N + M)\times (N + M)}$, a Euclidean distance matrix $\boldsymbol{D} \in \mathbb{R}^{(N + M)\times (N + M)}$.
    \State {\bfseries Output:} $K$ sampled points.
    \State $\tilde{\boldsymbol{A}} \leftarrow \sum_i \hat{a}_{ij} \in \mathbb{R}^{N+M}$ \hfill\Comment{sum up along rows}
    \State $\tilde{\boldsymbol{D}} \leftarrow \frac{\boldsymbol{D} - \min{\boldsymbol{D}}}{\max{\boldsymbol{D}} -\min{\boldsymbol{D}}}  \in \mathbb{R}^{(N + M)\times (N + M)}$ \hfill\Comment{normalize the distance matrix}
    \State $\mathcal{P} = \{\boldsymbol{x}_\#\}$
    \State $\mathcal{M} = \{\boldsymbol{x}_i\}_{i=1}^{N+M}$
    \While{${\rm length}(\mathcal{P}) < k$} 
    \State $\boldsymbol{x}_{\textrm{new}} \leftarrow \underset{i \in \mathcal{M}}{\mathrm{argmax}}\, (\underset{j \in \mathcal{P}}{\mathrm{min}}\, \tilde{D}_{ij} + \lambda \tilde{{A}}_{i})$ \hfill\Comment{pick up the node that maximize the objective}
    \State $\mathcal{P}$.append($\boldsymbol{x}_{\textrm{new}}$)
    \State $\mathcal{M}$.remove($\boldsymbol{x}_{\textrm{new}}$)
    \EndWhile
    \Return $\mathcal{P}$
\end{algorithmic}
\end{algorithm}

Specifically, we choose the virtual atom $\boldsymbol{x}_\#$ as the starting point and initialize two lists $\mathcal{P} =\{\boldsymbol{x}_\#\}$ and $\mathcal{M} =\{\boldsymbol{x}_i\}_{i=1}^{N+M}$ to store remaining candidate points. Then the process begins with the attention score matrix $\hat{\boldsymbol{A}}=[\hat{a}_{i,j}]_{i,j\in [N+M]}  \in \mathbb{R}^{(N + M)\times (N + M)}$ and the interatomic distance matrix $\boldsymbol{D} \in \mathbb{R}^{(N + M)\times (N + M)}$. It can be easily proved that each row of $\hat{\boldsymbol{A}}$ sums up to 1 after the $Softmax$ operation along columns, i.e. $\sum_j\hat{a}_{ij} = 1, \forall i\in [N+M]$. In order to obtain the importance of each atom in the self-attention computation, we accumulate $\hat{\boldsymbol{A}}$ along rows and get $\tilde{\boldsymbol{A}}=\sum_i \hat{a}_{ij}\in \mathbb{R}^{N+M}$. Besides, we adopt the min-max normalization to rescale the distance matrix $\boldsymbol{D}$ into values between 0 and 1, and obtain $\tilde{\boldsymbol{D}}=\frac{\boldsymbol{D} - \min{\boldsymbol{D}}}{\max\boldsymbol{D} - \min\boldsymbol{D}}$. 

After the above preprocess, we repeatedly move a point $\boldsymbol{x}_{\textrm{new}}$ from $\mathcal{M}$ to $\mathcal{P}$, which ensures that $\boldsymbol{x}_{\textrm{new}}$ is as far from $\mathcal{P}$ as possible by maximizing $\tilde{D}_{ij}$ and also plays a crucial role in attention computation by maximizing $\tilde{{A}}_{i} $. Mathematically, the AFPS aims to achieve the 
objective:
\begin{equation}
\label{equ: afps_object}
    \max \sum_{i \in \mathcal{M}} (\underset{j \in \mathcal{P} \setminus \{i\}}{\min} \tilde{D}_{ij} + \lambda \tilde{{A}}_{i}),
\end{equation}
where $\lambda$ is a hyperparameter to balance those two different goals. This process is repeated until $\mathcal{P}$ has reached $K$ points. Algorithm~\ref{agl: AFPS_algo} provides a greedy approximation solution to this AFPS optimization objective for sake of computational efficiency.

After that, sampled features $\{\boldsymbol{z}''_i\}_{i \in P}$ are gathered by a Global Average Pooling layer to attain the molecular representation $\boldsymbol{r}\in \mathbb{R}^{\psi_\textrm{model}}$.

Remarkably, our proposed AFPS has considerable differences and superiority over a body of previous hierarchical approaches~\citep{eismann2021hierarchical}. Their subsampling operations are mainly designed for protein complexities, which often have uniform structures. To be specific, they hierarchically use alpha carbons as the intermediate set of points and aggregate information at the level of those carbons for the entire complex. However, the structures of small molecules have no such stable paradigm, and we provide a universal method to adaptively subsample atoms without any prior assumptions on the atom arrangement. 

%% file: 5_experiment.tex
\section{Experiments}
We conduct extensive experiments on 7 datasets about both small molecules and large protein molecules from three different domains, including quantum chemistry, physiology, and biophysics. The appendix summarises statistical information of these 7 benchmark datasets, such as the number of tasks and task types, the number of molecules and atom classes, the minimum and the maximum number of atoms, and the density (mean interatomic distances) of all molecules. 

\paragraph{Datasets.} We test Molformer on a series of small molecule datasets, containing QM7~\citep{blum2009970}, QM8~\citep{ramakrishnan2015electronic}, QM9~\citep{ramakrishnan2014quantum}, BBBP~\citep{martins2012bayesian}, ClinTox~\citep{gayvert2016data}, and BACE~\citep{subramanian2016computational}~\footnote{For BBBP, ClinTox, and BACE, we use RDKit~\citep{landrum2013rdkit} to procure 3D coordinates from SMILES.}. QM7 is a subset of GDB-13 and is composed of 7K molecules. QM8 and QM9 are subsets of GDB-17 with 22K and 133K molecules respectively. BBBP involves records of whether a compound carries the permeability property of penetrating the blood-brain barrier. ClinTox compares drugs approved through FDA and drugs eliminated due to toxicity during clinical trials. BACE is collected for recording compounds that could act as the inhibitors of human $\beta$-secretase 1 (BACE-1).

We also inspect Molformer's ability to learn mutual relations between proteins and molecules on the PDBbind dataset~\citep{wang2005pdbbind}. Following \citet{townshend2020atom3d}, we split protein-ligand complexes by protein sequence identity at 30\%. As for the target, we predict $pS = -\log(S)$, where $S$ is the binding affinity in Molar unit. In addition, we only use the pocket of each protein and put pocket-ligand pairs together as the input.

For QM9, we use the exact train/validation/test split as~\citet{townshend2020atom3d}. For PDBbind, 90\% of the data is used for training and the rest is divided equally between validation and test like~\citet{chen2019graph}. For others, we adopt the scaffold splitting method with a ratio of 8:1:1 for train/validation/test as~\citet{rong2020self}. More implementing details can be found in Appendix.  % ~\ref{app: exp_detail}

\paragraph{Baselines}
For small molecules, we compare Molformer with the following baselines. TF\_Robust~\citep{ramsundar2015massively} takes molecular fingerprints as the input. %GraphConv~\citep{kipf2016semi}, 
Weave~\citep{kearnes2016molecular}, 
MPNN~\citep{gilmer2017neural}, Schnet~\citep{schutt2018schnet}, MEGNet~\citep{chen2019graph}, GROVER~\citep{rong2020self}, DMPNN~\citep{yang2019analyzing}, MGCN~\citep{lu2019molecular}, AttentiveFP~\citep{xiong2019pushing}, DimeNet~\citep{klicpera2020directional}, DimeNet++~\citep{klicpera2020fast}, PaiNN~\citep{schutt2021equivariant}, and SphereNet~\citep{liu2021spherical} are all graph convolutional models.  Graph Transformer~\citep{chen2019path}, MAT~\citep{maziarka2020molecule}, R-MAT~\citep{maziarka2021relative}, SE(3)-Transformer~\citep{fuchs2020se}, and LieTransformer~\citep{hutchinson2021lietransformer} are Transformer-based Equivariant Neural Networks (ENNs)~\citep{thomas2018tensor}. 

For PDBbind, we choose seven baselines. DeepDTA~\citep{ozturk2018deepdta} and DeepAffinity~\citep{karimi2019deepaffinity} take in pairs of ligand and protein SMILES as input. Cormorant~\citep{anderson2019cormorant} is an ENN that represents each atom by its absolute 3D coordinates. HoloProt~\citep{somnath2021multi} captures higher-level fingerprint motifs on the protein surface. Schnet, 3DCNN and 3DGCN~\citep{townshend2020atom3d} are 3D methods.

\begin{table}[t]
% \vspace{-2em}
\centering
\resizebox{1.0\columnwidth}{!}{%
\begin{tabular}{@{}l|rrrrr}
\toprule
Method  & QM7 & QM8 & BBBP & ClinTox & BACE \\ \midrule
TF-Robust~\citep{ramsundar2015massively} & 120.6 & 0.024 & 0.860 & 0.765 & 0.824\\ \midrule
%GraphConv~\citep{kipf2016semi} & 118.9 & 0.021 & 0.877 & 0.845 & 0.854 \\
Weave~\citep{kearnes2016molecular} & 94.7 & 0.022 & 0.837 & 0.823 & 0.791 \\
MPNN~\citep{gilmer2017neural} &  113.0 & 0.015 & 0.913 & 0.879  & 0.815 \\
Schnet~\citep{schutt2018schnet} &  74.2 & 0.020 & 0.847 & 0.717 & 0.750 \\
DMPNN~\citep{yang2019analyzing} & 105.8 & 0.014 & \underline{0.919} & 0.897 & 0.852 \\
MGCN~\citep{lu2019molecular} &  77.6 & 0.022 & 0.850 &  0.634 & 0.734 \\ 
Attentive FP~\citep{xiong2019pushing} & 126.7 & 0.028 & 0.908 & \underline{0.933}  & 0.863\\ \midrule
Graph Transformer~\citep{chen2019path} & \underline{47.8} & \underline{0.010} & 0.913 & - & \underline{0.880} \\
MAT~\citep{maziarka2020molecule} & 102.8 & - & 0.728 & - & 0.846 \\
R-MAT~\citep{maziarka2021relative} & 68.6 & - & 0.746 & - & 0.871 \\
GROVER\textsubscript{large}~\citep{rong2020self} &  89.4 & 0.017 & 0.911 & 0.884  & 0.858 \\
\midrule
Molformer &  \textbf{43.2} & \textbf{0.009} & \textbf{0.926} & \textbf{0.937} & \textbf{0.884} \\
\bottomrule
\end{tabular}
}
\caption{For regression tasks in QM7 and QM8, lower MAE is better. For classification tasks in BBBP, ClinTox, and Bace, higher values are better.}
% \vspace{-1em}
\label{tab: small_molecule}
\end{table}
\subsection{Overall Results on Benchmark Tasks}
\paragraph{Molecules.} Table~\ref{tab: small_molecule} and Table~\ref{tab: QM9} document the overall results on small molecules datasets, where the best performance is marked bold and the second best is underlined for clear comparison. It can be discovered that Molformer achieves the lowest MAE of 43.2 on QM7 and 0.009 on QM8, beating several strong baselines including DMPNN and Graph Transformer. Besides, Molformer offers competitive performance in all property regression tasks on QM9. Particularly, we outperform all Transformer-based ENNs, including SE(3)-Transformer and LieTransformer. As for classification problems, we surpass all baselines mostly by a fairly large margin.

\begin{table*}[ht]
\centering
\resizebox{2.0 \columnwidth}{!}{%
\begin{tabular}{@{} l|rrrrr rrrrrrr} \toprule
Target & $\epsilon$\textsubscript{HOMO} &  $\epsilon$\textsubscript{LUMO}  & $\Delta\epsilon$ & $\mu$ & $\alpha$ & $R^2$ &  ZPVE & $U_0$ & $U$ & $H$ & $G$ &  $c_v$ \\ 
Unit & eV & eV & eV & D & bohr\textsuperscript{3} & $a_0^2$ & meV & meV & meV & meV & meV & $\textrm{cal}/{\textrm{mol K}}$\\ \midrule
MPNN~\citep{gilmer2017neural} & .043 & .037 & .069 & .030 & .092 & .150 & 1.27 & 45 & 45 & 39 & 44 & .800  \\
Schnet~\citep{schutt2018schnet} & .041 & .034 & .063 & .033 & .235 & \underline{.073} & 1.7 & 14 & 19 & 14 & 14 & .033  \\
MEGNet~\textsubscript{full}~\citep{chen2019graph} &  .038 & .031 & .061 & .040 & .083 & .265 & 1.4 & 9 & 10 & 10 & 10 & .030   \\ 
MGCN~\citep{lu2019molecular} & .042 & .057 & .064 & .056 & \textbf{.030} & .110 & \textbf{1.12} & 12.9 & 14.4 & 14.6 & 16.2 & .038 \\
DimeNet~\citep{klicpera2020directional} & .027 & .019 & .034 & .028 & .046  & .331 & 1.29 & 8.02 & 7.89 & 8.11 & 8.98 & .024 \\ 
DimeNet++~\citep{klicpera2020fast} & \underline{.024} & \underline{.019} & \underline{.032} & .029  & \underline{.043}  & .331 & 1.21 & 6.32 & \underline{6.28} & 6.53 & \underline{7.56} & \underline{.023} \\
SphereNet~\citep{liu2021spherical} & \textbf{.024} & \textbf{.019} & \textbf{.032} & \underline{.026}  & .047 & .292 & \underline{1.12} & \underline{6.26} & 7.33 & \underline{6.40} & 8.0 & \textbf{.021} \\
PaiNN~\citep{schutt2021equivariant} & .028 & .020 & .046 & \textbf{.012} & .045 & \textbf{.066} & 1.28 & \textbf{5.85} & \textbf{5.53} & \textbf{5.98} & \textbf{7.35} & .024  \\ \midrule
SE(3)-Transformer~\citep{fuchs2020se} & .035 & .033 & .053 & .051 & .142 & --& --& --& --& --& --& -- \\
LieTransformer~\citep{hutchinson2021lietransformer} &  .033 & .029 & .052 & .061 & .104 & 2.29 & 3.55 & 17 & 16 & 27 & 23  & .041 \\ \midrule
Molformer & {.025} & .026  & .039  & .028 & .041 & .350 & 2.05 & 7.52 & 7.46 & 7.38 & 8.11 & .025 \\ \bottomrule
\end{tabular}
}
\caption{Comparison of MAE on QM9. The last three methods are Transformer-based methods.}
\label{tab: QM9}
\end{table*}

\begin{table}[t]
% \vspace{-1em}
\centering
\resizebox{1.0\columnwidth}{!}{%
\begin{tabular}{@{} l|l|rrr @{}} \toprule
Method & Geometry & RMSD & $R_p$ &  $R_s$ \\ \midrule
DeepDTA~\citep{ozturk2018deepdta} & Non-3D & 1.565 & 0.573 & 0.574 \\
DeepAffinity~\citep{karimi2019deepaffinity} & Non-3D & 1.893 & 0.415 & 0.426 \\ \midrule
Schnet~\citep{schutt2018schnet} & 3D & 1.892 &\underline{0.601} & - \\
Cormorant~\citep{anderson2019cormorant} & 3D & \underline{1.429} & 0.541 & 0.532 \\
3DCNN~\citep{townshend2020atom3d} & 3D & 1.520 & 0.558 & 0.556 \\
3DGCN~\citep{townshend2020atom3d} & 3D & 1.963 & 0.581 & \underline{0.647} \\ 
HoloProt~\citep{somnath2021multi} & 3D & 1.464 & 0.509 & 0.500 \\ \midrule
Molformer & 3D & \textbf{1.386} & \textbf{0.623} & \textbf{0.651} \\ \bottomrule
\end{tabular}
}
\caption{Comparison of RMSD, $R_p$, and $R_s$ on PDBbind.}
\label{tab: LBA}
\end{table}
\paragraph{Proteins.} Table~\ref{tab: LBA} reports the Root-Mean-Squared Deviation (RMSD), the Pearson correlation ($R_p$), and the Spearman correlation ($R_s$) on PDBbind. Molformer achieves the lowest RMSD and the best Pearson and Spearman correlations. As~\citet{wu2018moleculenet} claim, appropriate featurization which holds pertinent information is significant for PDBbind. However, an important observation in our work is that deep learning approaches with the exploitation of 3D geometric information can perform better than conventional methods like DeepDTA and DeepAffinity which use a set of physicochemical descriptors but ignore 3D structures.  

\subsection{Ablation Study and Discussion}
\begin{table}[t]
% \vspace{-1em}
\centering
\resizebox{0.8\columnwidth}{!}{
\begin{tabular}{@{}c|ccc|cccc@{}}
\toprule
 & HSA & AFPS & HMG & QM7 & QM8  &  PDBbind \\ \midrule
1 & - & -  & - & 132.2 & 0.0205 & 1.925 \\
2 & \checkmark & -  & \checkmark & 46.5 & \underline{0.0097} &  1.441  \\
3 & \checkmark & \checkmark & \checkmark & \textbf{43.2} & \textbf{0.0095} &  \textbf{1.386} \\ \bottomrule
\end{tabular}
}
\caption{Effects of each module on QM7, QM8 and PDBbind (RMSD).}
\label{tab: ablation}
\end{table}
\paragraph{What Is the Effect of Each Component?} We investigate the effectiveness of each component of our Molformer in Table~\ref{tab: ablation}. It can be observed that HSA along with HMGs substantially boosts the model's performance compared with the naive method that immediately adds 3D coordinates as the atom input feature. MAE declines from 132.2 to 46.5 in QM7 while decreasing from 0.0205 to 0.0097 in QM8. In addition, AFPS produces better predictions than the counterpart that utilizes the virtual node as the molecular representation (a case study of AFPS is in the Appendix). We also discover that the multi-scale mechanism significantly reduces RMSD from 50.1 to 46.5 on QM7, but its improvements in QM8 are much smaller. This phenomenon indicates that the multi-scale mechanism is an appropriate way to alleviate the problem of inadequate training in small datasets. It endows Molformer with the capability to extract local features by regulating the scope of self-attention. However, as the data size gets larger, Molformer does not require the assistance of the multi-scale mechanism to abstract local patterns, since the parameters of convolution operators are properly trained. 

\begin{table}[t]
% \vspace{-1em}
\centering
\resizebox{0.75\columnwidth}{!}{
\begin{tabular}{@{}c|ccc}
\toprule
 & QM7 & QM8  &  PDBbind \\ \midrule
No Motifs & 132.2 & 0.0205 &  1.925 \\
Multi-Level Fusion & \underline{89.7} & \underline{0.0154} & \underline{1.427} \\
Heterogeneous Graphs &  \textbf{43.2} & \textbf{0.0095} &  \textbf{1.386} \\
\bottomrule
\end{tabular}}
\caption{Comparison of HMGs with simple feature fusion on QM7, QM8 and PDBbind (RMSD).}
\label{tab: multi_level}
\end{table}
\paragraph{What Is the Contribution of HMGs?}
\label{motif_role}
The ideology of constructing heterogeneous graphs has already been proven successful in not only chemical knowledge graphs, but named entity recognition~\citep{li2020flat}. The former views the chemical characteristics obtained from domain knowledge of elements as shared nodes, while the latter converts the lattice structure into a flat structure consisting of spans. To further verify its efficacy, we compare our motif-based HMGs with the naive fusion of multi-level features. Table~\ref{tab: multi_level} shows a noticeable improvement in our HMGs over the other two variants.

\begin{figure}[t]
\centering
\includegraphics[scale=0.25]{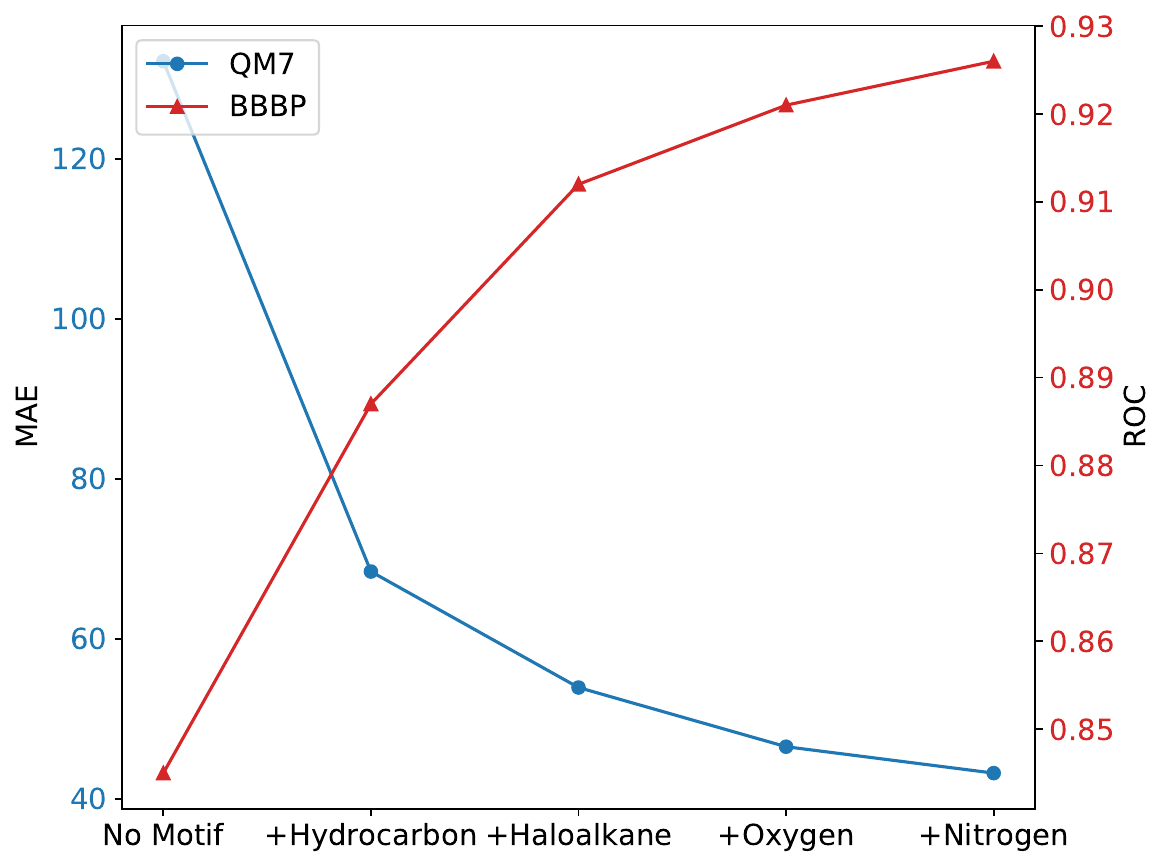}
\caption{The ablation study of consecutively adding four motif groups (from left to right) in the QM7 and BBBP datasets.}
\label{fig: ablation_motif}
% \vspace{-1em}
\end{figure}
\paragraph{Have We Found Good Candidates of Motifs?}
\label{motif_groups}
How to determine motifs is critical to HMGs. Concerning small molecules, we define motifs on the basis of functional groups, which refers to a substituent or moiety that causes molecules' characteristic chemical reactions. To further explore their contributions, we divide functional groups into four categories: Hydrocarbons, Haloalkanes, groups that contain oxygen, and groups that contain nitrogen (see Appendix). The ablation studies (see Figure~\ref{fig: ablation_motif}) demonstrate that Molformer can gain improvements from all four groups of motifs, where Hydrocarbons and Haloalkanes are the most and the least effective types respectively. This is in line with the fact that Hydrocarbons occur most frequently in organic molecules. Moreover, the best performance is achieved when all categories are considered, implying a promising direction to discover more effective motifs.
As for proteins, motifs discovered by our RL mining method share the same backbone as  $\emph{CC(C(NC(C)C(O)=O)=O)NC(CNC(CN)=O)=O}$ (see Appendix), which is a hydrogen bond donor and implies a mark to distinguish potential binding site. Moreover, the portion of those motifs in the pocket ($1.38\%$) is nearly twice that in other locations ($0.73\%$), conforming to the fact that pockets are the most preferable part for ligands to bind with. 
%We offer additional experiments of three-amino-acid motif in Appendix~\ref{rl_motif_additional}. 

%% file: 6_related_work.tex
\section{Related Works}
\paragraph{Motifs in Molecular Graphs.} Motifs have been proven to benefit many tasks from exploratory analysis to transfer learning. Various algorithms have been proposed to exploit them for contrastive learning~\citep{zhang2020motif}, self-supervised pretraining~\citep{zhang2021motif}, generation~\citep{jin2020hierarchical}, protein design~\citep{literminator} and drug-drug interaction prediction~\citep{huang2020caster}. To the best of our knowledge, none of them take advantage of motifs to build a heterogeneous graph for molecular property prediction. 

As for motif extraction, previous motif mining methods either depend on exact counting~\citep{cantoni20113d} or sampling and statistical estimation~\citep{wernicke2006efficient}. No preceding studies extract task-specific motifs to enhance model performance. 

\paragraph{Molecular Representation Learning.} DL has been widely applied to predict molecular properties. Molecules are usually represented as 1D sequences, including amino acid sequences and SMILES~\citep{xu2017seq2seq}, and 2D graphs~\citep{duvenaud2015convolutional}. Despite that, more evidence indicates that 3D structures lead to better modeling and superior performance. 3D CNNs~\citep{anand2021protein} and GNNs~\citep{cho2018three} become popular to capture these complex geometries in various bio-molecular applications. Nonetheless, the aforementioned methods is inefficient at grabbing local contextual feature and long-range dependencies.

Attempts have been taken to address that issue based on the Transformer. It assumes full connection and uses self-attention to capture long-term dependencies. Some researchers feed SMILES in Transformer to obtain representations~\citep{honda2019smiles} and conduct pretraining~\citep{chithrananda2020chemberta}. 
Others employ Transformer to solve generative tasks~\citep{ingraham2019generative} or fulfill equivariance~\citep{fuchs2020se} via spherical harmonics. However, the foregoing methods are either incapable to encode 3D geometry, non-sensitive to local contextual patterns, or inefficient to aggregate atom features. More essentially, they are not specially designed to operate on heterogeneous graphs of molecules. 

%% file: 7_conclusion.tex
\section{Conclusion}
This paper presents Molformer for 3D molecular representation learning on heterogeneous molecular graphs. First, we extract informative motifs by means of functional groups and a reinforcement learning mining method to formulate heterogeneous molecular graphs. After that, Molformer adopts a heterogeneous self-attention to distinguish the interactions between multi-level nodes and exploit spatial information with multiplicate scales for the sake of catching local features. Then a simple but efficient downsampling algorithm is introduced to better accumulate molecular representations. Experiments demonstrate the superiority of Molformer in various domains, which beats a group of baseline algorithms.

%% file: 8_appendix.tex
\section{Experimental Setup}
\label{app: exp_detail}
\subsection{Molformer Architecture}
A standard Molformer has 6 heterogeneous self-attention layers, and each layer has 3 scales and 8 heads. Normally, scales are set by $\tau=[\frac{\rho}{2}, \rho, 2\rho]$, where $\rho$ is the density of each corresponding dataset. The number of selected atoms $K$ and the weight ratio $\lambda$ in AFPS is set as 4 and 0.1, respectively. We use ReLU as the activation function and a dropout rate of 0.1 for all layers if not specified. The input embedding size is 512 and the hidden size for FFN is 2048. 
For BBBP and ClinTox, we use Molformer with 2 heterogeneous self-attention layers with 4 heads. The scales are 0.8, 1.6, and 3.0 $\textup{\AA}$. The dropout rate is 0.2 and 0.6 for BBBP and ClinTox, respectively. For BACE, we use a standard Molformer but with a dropout rate of 0.2.

\subsection{Training Details}
We use Pytorch to implement Molformer and data parallelism in two GeForce RTX 3090. An Adam optimizer is used and a ReduceLROnPlateau scheduler is enforced to adjust it with a factor of 0.6 and a patience of 10. We apply no weight decay there. Each model is trained with 300 epochs, except for PDBbind where we solely train the model for 30 epochs. The data summary is in Table~\ref{tab: Datasets}.

For QM7 and QM8, we use a batch size of 128 and a learning rate of $10^{-4}$. For QM9, we use a batch size of 256 and a learning rate of $10^{-3}$. For PDBbind, we use a batch size of 16 and a learning rate of $10^{-4}$. All hyper-parameters are tuned based on validation sets. For all molecular datasets, we impose no limitation on the input length and normalize the values of each regression task by mean and the standard deviation of the training set. We used grid search to tune the hyper-parameters of our model and baselines based on the validation dataset (see Table~\ref{tab: hyper}).
\begin{table*}[h]
% \vspace{-1em}
\centering
\resizebox{2.0\columnwidth}{!}{
\begin{tabular}{@{}c|c|c}
\toprule
Hyper-parameter & Description & Range \\ \midrule
bs &  The input batch size. & [16, 128, 256]   \\
lr & The initial learning rate of ReduceLROnPlateau learning rate scheduler. & [$1e-4$, $1e-5$] \\
min\_lr &  The minimum learning rate of ReduceLROnPlateau learning rate scheduler. & $5e-7$ \\
dropout & The dropout ratio. & [0.1, 0.2, 0.6] \\
n\_encoder & The number of heterogeneous self-attention layers. &  [2, 6] \\
head & The number of self-attention heads. & [4, 8] \\ 
embed\_dim & The dimension of input embeddings. & 512 \\
ffn\_dim & The hidden size of MLP layers. & 1024 \\
k & The number of sampled points in AFPS. & [4, 8, 10, 20]\\ 
lambda & The balance ratio in AFPS. & [0.1, 0.5, 1.0, 2.0] \\ 
dist\_bar & The scales of the multi-scale mechanism (\textup{\AA}). & [[0.75, 1.55, 3.0], [0.8, 1.6, 3.0], [1.5, 3.0, 6.0]]\\
\bottomrule
\end{tabular}}
\caption{The training hyper-parameters.}
\label{tab: hyper}
\end{table*}

\begin{table*}[h]
\centering
\resizebox{1.8\columnwidth}{!}{%
\begin{tabular}{@{}c| ccccccccc}
\toprule
\textbf{Category} & \textbf{Dataset} & \textbf{Tasks} & \textbf{Task Type} & \textbf{Molecules} & \textbf{\begin{tabular}[c]{@{}c@{}}Atom\\  Class\end{tabular}} & \textbf{\begin{tabular}[c]{@{}c@{}}Min. \\ Atoms\end{tabular}} & \textbf{\begin{tabular}[c]{@{}c@{}}Max. \\ Atoms\end{tabular}} & \textbf{\begin{tabular}[c]{@{}c@{}}Density \\ (\textup{\AA})\end{tabular}}  & \textbf{\begin{tabular}[c]{@{}c@{}}Metric \\ \end{tabular}} \\ \midrule
\multirow{3}{*}{\begin{tabular}[c]{@{}c@{}}Quantum \\ Chemistry\end{tabular}} & QM7 & 1 & regression & 7,160 & 5 & 4 & 23 & 2.91  & MAE \\
 & QM8 & 12 & regression & 21,786 & 5 & 3 & 26 & 1.54 & MAE \\
 & QM9 & 12 & regression & 133,885 & 5 & 3 & 28 & 1.61 & MAE\\ \midrule
\multirow{2}{*}{\begin{tabular}[c]{@{}c@{}}Physiology \end{tabular}} & BBBP & 1 & classification & 2,039 & 13 & 2 & 132 & 2.64  & ROC-AUC\\
 & ClinTox & 2 & classification & 1,478 & 27 & 1 & 136 & 2.83 & ROC-AUC\\  \midrule
\multirow{2}{*}{\begin{tabular}[c]{@{}c@{}}Biophysics \end{tabular}}
& PDBind\tablefootnote{The total number of proteins in the full, unsplit PDBbind is 11K, but our experiment only uses 4K proteins at 30\% sequence identity. Moreover, the number of atoms is the sum of both the pocket and molecules.} & 1 & regression & 11,908 & 23 & 115 & 1,085 & 5.89 & RMSE \\ 
& BACE & 1 & classification & 1,513 & 8 & 10 & 73 & 3.24 & ROC-AUC \\
\bottomrule
\end{tabular}}
\caption{Key statistics of datasets from three different categories.}
\label{tab: Datasets}
\end{table*}

\subsection{Motif Extraction}
\label{app: motif_gene}
\begin{figure*}[h]
\centering
\includegraphics[scale=0.4]{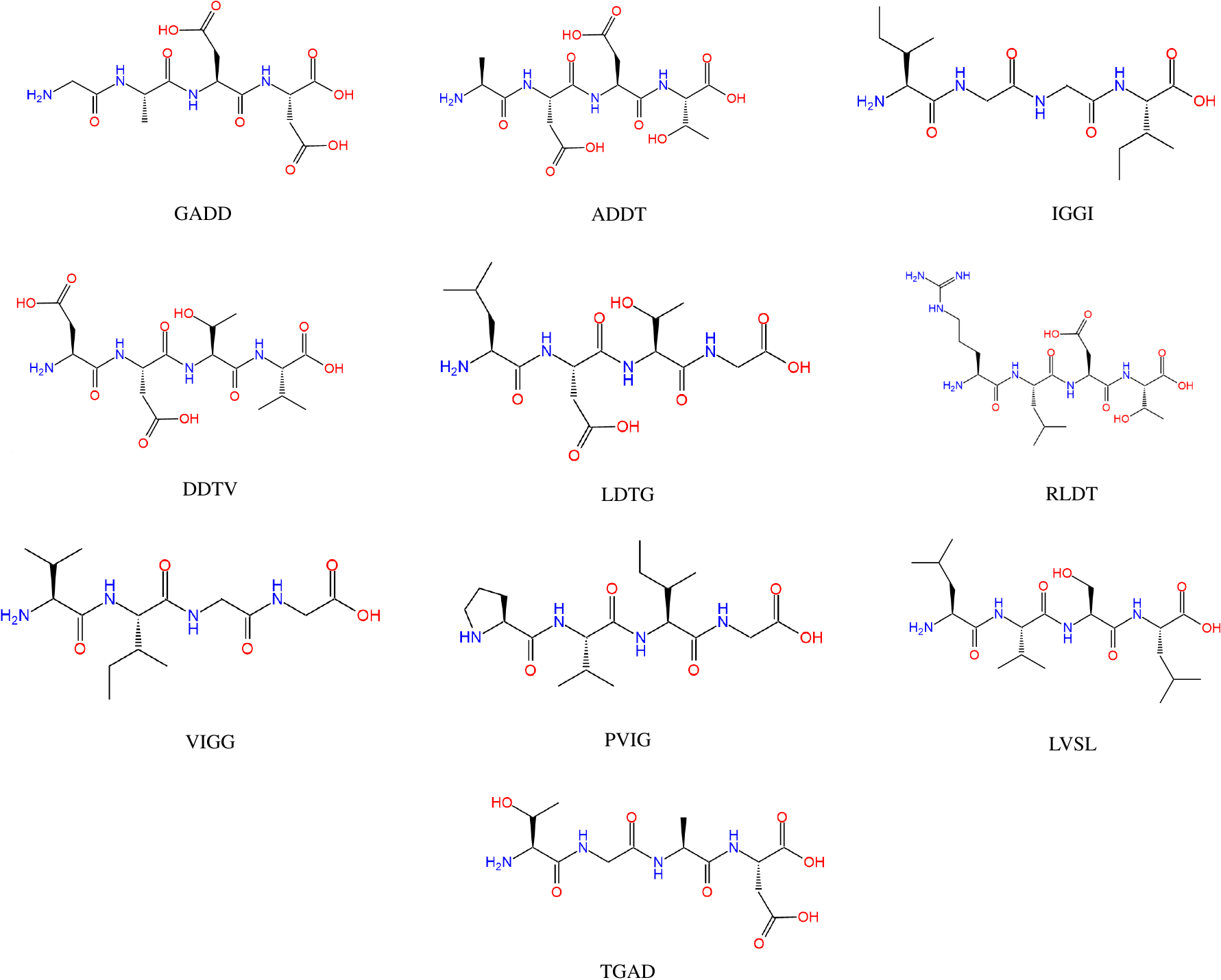}
\caption{\textbf{The motif lexicon found by our RL method,} where each motif is composed of ten quaternary amino acids and the upper-case names correspond to their compositions.}
\label{fig: protein_motif}
\end{figure*}

We adopt RDKit to search motifs from SMILES representations of small molecules. For QM9 and PDBbind, Atom3D provides both 3D coordinates and SMILES. For QM7, DeepChem offers SMILES. For QM8, we first attain SMILES from their 3D representations using RDKit and then extract motifs.
\begin{figure*}[h]
\centering
\includegraphics[scale=0.65]{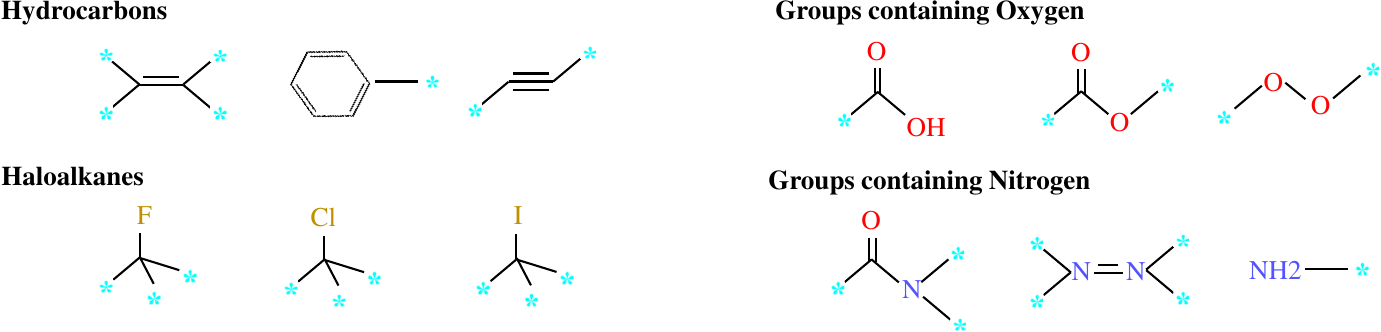}
\caption{\textbf{Examples of the four different motif categories} that we apply in Molformer based on functional groups.}
\label{fig: motif}
% \vspace{-1em}
\end{figure*}
For motifs in proteins,   Besides, we utilize a Latin hyper-cube sampling to sample 1K quaternary amino acids as the candidates in each iteration, and $\gamma$ is set as $1e^{-3}$. The motif lexicon explored by our RL method for the PDBbind task is in Figure~\ref{fig: protein_motif}. 

For motifs in proteins, since we aim to mine the optimal task-specific lexicon, it is unnecessary to take into account all 160,000 possible quaternions. Instead, we only need to consider the existing quaternions in the datasets. Specifically, there are only 29,871 kinds of quaternions in PDBbind. Besides, we utilize a Latin hyper-cube sampling to sample 1K quaternary amino acids as the candidates in each iteration, and $\gamma$ is set as $1e^{-3}$. The motif lexicon explored by our RL method for the PDBbind task is in Figure~\ref{fig: protein_motif}. 
%However, due to the limitation of GPU memory, it is still impossible to store the gradients of all 29,871 candidate quaternions for backpropagations.
Moreover, the portion of motifs in a protein (or some part such as a pocket) is the number of motifs divided by the number of all amino acids in that protein.

\section{Additional Experimental Results}
\subsection{Conformation Classification}
\label{classification}
\paragraph{Task and Data.} To explore the influence of multiple conformations, we introduce a new task, conformation classification, to evaluate the model's capacity to differentiate molecules with various low-energy conformations. We use the recent GEOM-QM9 which is an extension to the QM9 dataset. It contains multiple conformations for most molecules, while the original QM9 only contains one. 

We randomly draw 1000 different molecules from GEOM-QM9, each with 20 different conformations. Models are required to distinguish the molecular type given different conformations. We take half of each molecule's conformations as the training set and another half as the test split. Since it is a multi-class classification problem with 1000 classes, we compute the micro-average and macro-average ROC-AUC as well as the accuracy for evaluations. 

\paragraph{Results.} Molformer achieves a perfect micro-average and macro-average ROC-AUC  as well as a high accuracy (see Table~\ref{tab: conformation}). This indicates the strong robustness of our model against different spatial conformations of molecules. 
\begin{table*}[h]
% \vspace{-1em}
\centering
\resizebox{0.6\columnwidth}{!}{
\begin{tabular}{@{} c| ccc}
\toprule
Metrics & Acc. & Micro. & Macro.\\ \midrule
Molformer & 0.999 & 1.000 & 1.000 \\
\bottomrule
\end{tabular}
}
\caption{Molformer performance on conformation classification.}
\label{tab: conformation}
% \vspace{-2em}
\end{table*}

\subsection{AFPS vs. FPS.}
\label{case_afps}
\begin{figure*}[ht]
% \vspace{-0.5em}
\centering
\includegraphics[scale=0.25]{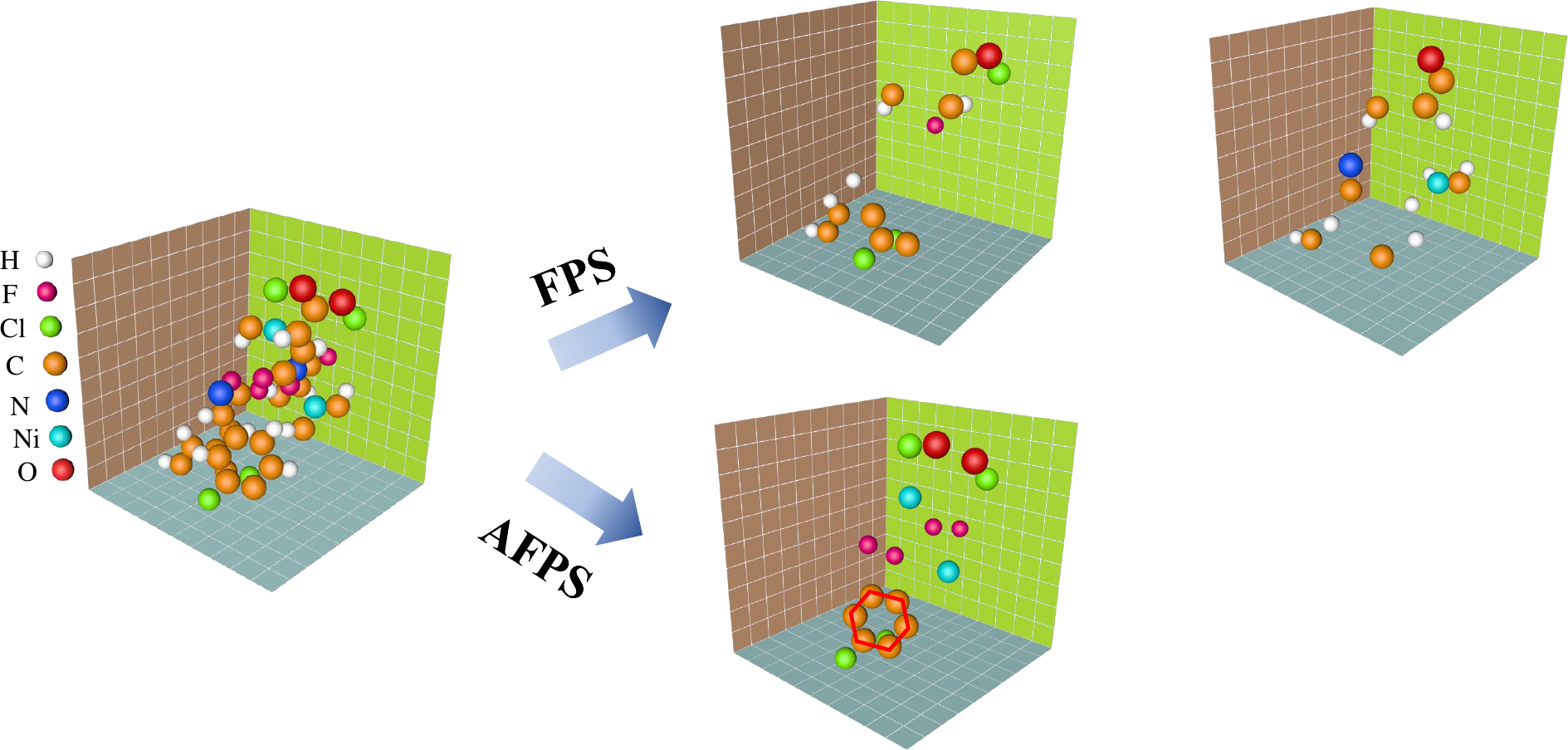}
\caption{Sampled points using FPS and AFPS. The red circle represents a benzene ring. We do not show dummy nodes there.}
% \vspace{-1.5em}
\label{fig: downsample}
\end{figure*}

To have a vivid understanding of the atom sampling algorithm, we conducted a case study on a random molecule (see Figure~\ref{fig: downsample}). Points selected by FPS are randomized and exclude vital atoms like the heavy metal Nickel (Ni). With the adoption of AFPS, sampled points include Ni, Nitrogen (N), and the benzene ring besides that they keep remote distances from each other. Moreover, FPS integrates too many features of trivial atoms like Hydrogen (H) while missing out on key atoms and motifs, which will significantly smooth the molecular representations and lead to poor predictions. This illustrative example shows the effectiveness of our AFPS to offset the disadvantages of the conventional FPS in 3D molecular representation. 

\subsection{Protein}
\begin{figure*}[h]
\centering
\includegraphics[scale=0.35]{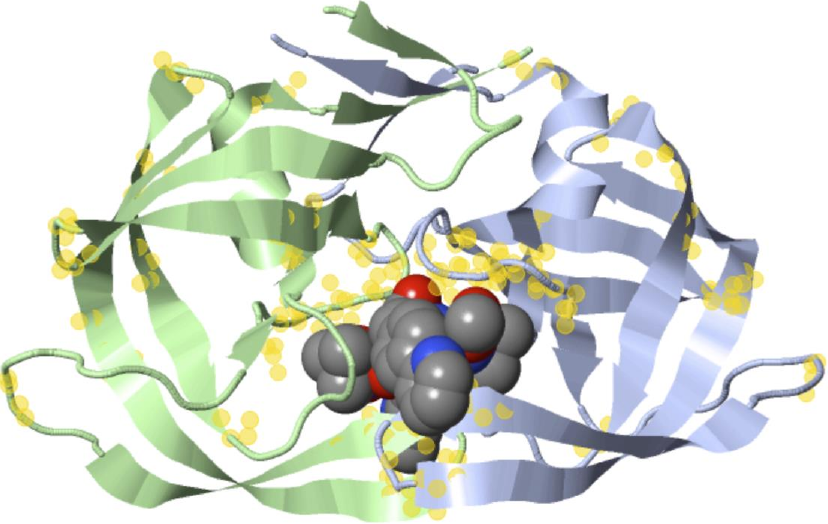}
\caption{The protein-ligand pair of 2aqu in PDBbind. The yellow dot-halos denotes the motifs found in this protein.}
\label{fig: protein_ligand}
\end{figure*}

We envision a protein-ligand pair in PDBbind in Figure~\ref{fig: protein_ligand}. It can be observed that motifs occur much more frequently in the area of the protein pocket than in other places. To be specific, the ligand is exactly surrounded by our discovered motifs, which strongly demonstrates the effectiveness of our RL method to mine motifs with semantic meanings. 
% \subsection{RL Motif Mining for Three-Amino-Acid Motifs}
% \label{rl_motif_additional}
% To further explore what characteristic those motifs extracted by RL jointly have, we implement an auxiliary experiment, where the motif consists of only three amino acids. Thus, there are $20^3=8000$ potential motifs and $C_{8000}^{K}$ lexicons in $\mathbb{V}$. We keep other settings unchanged, and still find the optimal vocabulary with $K=10$ motifs. 

%% file: main.bbl
\begin{thebibliography}{59}
\providecommand{\natexlab}[1]{#1}

\bibitem[{Anand-Achim et~al.(2021)Anand-Achim, Eguchi, Mathews, Perez, Derry,
  Altman, and Huang}]{anand2021protein}
Anand-Achim, N.; Eguchi, R.~R.; Mathews, I.~I.; Perez, C.~P.; Derry, A.;
  Altman, R.~B.; and Huang, P. 2021.
\newblock Protein sequence design with a learned potential.
\newblock \emph{bioRxiv}, 2020--01.

\bibitem[{Anderson, Hy, and Kondor(2019)}]{anderson2019cormorant}
Anderson, B.; Hy, T.-S.; and Kondor, R. 2019.
\newblock Cormorant: Covariant molecular neural networks.
\newblock \emph{arXiv preprint arXiv:1906.04015}.

\bibitem[{Blum and Reymond(2009)}]{blum2009970}
Blum, L.~C.; and Reymond, J.-L. 2009.
\newblock 970 million druglike small molecules for virtual screening in the
  chemical universe database GDB-13.
\newblock \emph{Journal of the American Chemical Society}, 131(25): 8732--8733.

\bibitem[{Cantoni, Gatti, and Lombardi(2011)}]{cantoni20113d}
Cantoni, V.; Gatti, R.; and Lombardi, L. 2011.
\newblock 3D Protein Surface Segmentation through Mathematical Morphology.
\newblock In \emph{International Joint Conference on Biomedical Engineering
  Systems and Technologies}, 97--109. Springer.

\bibitem[{Chen, Barzilay, and Jaakkola(2019)}]{chen2019path}
Chen, B.; Barzilay, R.; and Jaakkola, T. 2019.
\newblock Path-augmented graph transformer network.
\newblock \emph{arXiv preprint arXiv:1905.12712}.

\bibitem[{Chen et~al.(2019)Chen, Ye, Zuo, Zheng, and Ong}]{chen2019graph}
Chen, C.; Ye, W.; Zuo, Y.; Zheng, C.; and Ong, S.~P. 2019.
\newblock Graph networks as a universal machine learning framework for
  molecules and crystals.
\newblock \emph{Chemistry of Materials}, 31(9): 3564--3572.

\bibitem[{Chithrananda, Grand, and Ramsundar(2020)}]{chithrananda2020chemberta}
Chithrananda, S.; Grand, G.; and Ramsundar, B. 2020.
\newblock Chemberta: Large-scale self-supervised pretraining for molecular
  property prediction.
\newblock \emph{arXiv preprint arXiv:2010.09885}.

\bibitem[{Cho and Choi(2018)}]{cho2018three}
Cho, H.; and Choi, I.~S. 2018.
\newblock Three-dimensionally embedded graph convolutional network (3dgcn) for
  molecule interpretation.
\newblock \emph{arXiv preprint arXiv:1811.09794}.

\bibitem[{Duvenaud et~al.(2015)Duvenaud, Maclaurin, Aguilera-Iparraguirre,
  G{\'o}mez-Bombarelli, Hirzel, Aspuru-Guzik, and
  Adams}]{duvenaud2015convolutional}
Duvenaud, D.; Maclaurin, D.; Aguilera-Iparraguirre, J.; G{\'o}mez-Bombarelli,
  R.; Hirzel, T.; Aspuru-Guzik, A.; and Adams, R.~P. 2015.
\newblock Convolutional networks on graphs for learning molecular fingerprints.
\newblock \emph{arXiv preprint arXiv:1509.09292}.

\bibitem[{Eismann et~al.(2021)Eismann, Townshend, Thomas, Jagota, Jing, and
  Dror}]{eismann2021hierarchical}
Eismann, S.; Townshend, R.~J.; Thomas, N.; Jagota, M.; Jing, B.; and Dror,
  R.~O. 2021.
\newblock Hierarchical, rotation-equivariant neural networks to select
  structural models of protein complexes.
\newblock \emph{Proteins: Structure, Function, and Bioinformatics}, 89(5):
  493--501.

\bibitem[{Elnaggar et~al.(2020)Elnaggar, Heinzinger, Dallago, Rihawi, Wang,
  Jones, Gibbs, Feher, Angerer, Steinegger et~al.}]{elnaggar2020prottrans}
Elnaggar, A.; Heinzinger, M.; Dallago, C.; Rihawi, G.; Wang, Y.; Jones, L.;
  Gibbs, T.; Feher, T.; Angerer, C.; Steinegger, M.; et~al. 2020.
\newblock ProtTrans: towards cracking the language of Life's code through
  self-supervised deep learning and high performance computing.
\newblock \emph{arXiv preprint arXiv:2007.06225}.

\bibitem[{Fuchs et~al.(2020)Fuchs, Worrall, Fischer, and Welling}]{fuchs2020se}
Fuchs, F.~B.; Worrall, D.~E.; Fischer, V.; and Welling, M. 2020.
\newblock Se (3)-transformers: 3d roto-translation equivariant attention
  networks.
\newblock \emph{arXiv preprint arXiv:2006.10503}.

\bibitem[{Gayvert, Madhukar, and Elemento(2016)}]{gayvert2016data}
Gayvert, K.~M.; Madhukar, N.~S.; and Elemento, O. 2016.
\newblock A data-driven approach to predicting successes and failures of
  clinical trials.
\newblock \emph{Cell chemical biology}, 23(10): 1294--1301.

\bibitem[{Gilmer et~al.(2017)Gilmer, Schoenholz, Riley, Vinyals, and
  Dahl}]{gilmer2017neural}
Gilmer, J.; Schoenholz, S.~S.; Riley, P.~F.; Vinyals, O.; and Dahl, G.~E. 2017.
\newblock Neural message passing for quantum chemistry.
\newblock In \emph{International conference on machine learning}, 1263--1272.
  PMLR.

\bibitem[{Guo et~al.(2020)Guo, Qiu, Liu, Xue, and Zhang}]{guo2020multi}
Guo, Q.; Qiu, X.; Liu, P.; Xue, X.; and Zhang, Z. 2020.
\newblock Multi-scale self-attention for text classification.
\newblock In \emph{Proceedings of the AAAI Conference on Artificial
  Intelligence}, volume~34, 7847--7854.

\bibitem[{Honda, Shi, and Ueda(2019)}]{honda2019smiles}
Honda, S.; Shi, S.; and Ueda, H.~R. 2019.
\newblock Smiles transformer: Pre-trained molecular fingerprint for low data
  drug discovery.
\newblock \emph{arXiv preprint arXiv:1911.04738}.

\bibitem[{Huang et~al.(2020)Huang, Xiao, Hoang, Glass, and
  Sun}]{huang2020caster}
Huang, K.; Xiao, C.; Hoang, T.; Glass, L.; and Sun, J. 2020.
\newblock Caster: Predicting drug interactions with chemical substructure
  representation.
\newblock In \emph{Proceedings of the AAAI Conference on Artificial
  Intelligence}, volume~34, 702--709.

\bibitem[{Hutchinson et~al.(2021)Hutchinson, Le~Lan, Zaidi, Dupont, Teh, and
  Kim}]{hutchinson2021lietransformer}
Hutchinson, M.~J.; Le~Lan, C.; Zaidi, S.; Dupont, E.; Teh, Y.~W.; and Kim, H.
  2021.
\newblock Lietransformer: Equivariant self-attention for lie groups.
\newblock In \emph{International Conference on Machine Learning}, 4533--4543.
  PMLR.

\bibitem[{Ingraham et~al.(2019)Ingraham, Garg, Barzilay, and
  Jaakkola}]{ingraham2019generative}
Ingraham, J.; Garg, V.~K.; Barzilay, R.; and Jaakkola, T. 2019.
\newblock Generative models for graph-based protein design.
\newblock \emph{Advances in neural information processing systems}.

\bibitem[{Ishida et~al.(2021)Ishida, Miyazaki, Sugaya, and
  Omachi}]{ishida2021graph}
Ishida, S.; Miyazaki, T.; Sugaya, Y.; and Omachi, S. 2021.
\newblock Graph Neural Networks with Multiple Feature Extraction Paths for
  Chemical Property Estimation.
\newblock \emph{Molecules}, 26(11): 3125.

\bibitem[{Jin, Barzilay, and Jaakkola(2020)}]{jin2020hierarchical}
Jin, W.; Barzilay, R.; and Jaakkola, T. 2020.
\newblock Hierarchical generation of molecular graphs using structural motifs.
\newblock In \emph{International Conference on Machine Learning}, 4839--4848.
  PMLR.

\bibitem[{Jumper et~al.(2021)Jumper, Evans, Pritzel, Green, Figurnov,
  Ronneberger, Tunyasuvunakool, Bates, {\v{Z}}{\'\i}dek, Potapenko
  et~al.}]{jumper2021highly}
Jumper, J.; Evans, R.; Pritzel, A.; Green, T.; Figurnov, M.; Ronneberger, O.;
  Tunyasuvunakool, K.; Bates, R.; {\v{Z}}{\'\i}dek, A.; Potapenko, A.; et~al.
  2021.
\newblock Highly accurate protein structure prediction with AlphaFold.
\newblock \emph{Nature}, 596(7873): 583--589.

\bibitem[{Karimi et~al.(2019)Karimi, Wu, Wang, and
  Shen}]{karimi2019deepaffinity}
Karimi, M.; Wu, D.; Wang, Z.; and Shen, Y. 2019.
\newblock DeepAffinity: interpretable deep learning of compound--protein
  affinity through unified recurrent and convolutional neural networks.
\newblock \emph{Bioinformatics}, 35(18): 3329--3338.

\bibitem[{Kearnes et~al.(2016)Kearnes, McCloskey, Berndl, Pande, and
  Riley}]{kearnes2016molecular}
Kearnes, S.; McCloskey, K.; Berndl, M.; Pande, V.; and Riley, P. 2016.
\newblock Molecular graph convolutions: moving beyond fingerprints.
\newblock \emph{Journal of computer-aided molecular design}, 30(8): 595--608.

\bibitem[{Klicpera et~al.(2020)Klicpera, Giri, Margraf, and
  G{\"u}nnemann}]{klicpera2020fast}
Klicpera, J.; Giri, S.; Margraf, J.~T.; and G{\"u}nnemann, S. 2020.
\newblock Fast and uncertainty-aware directional message passing for
  non-equilibrium molecules.
\newblock \emph{arXiv preprint arXiv:2011.14115}.

\bibitem[{Klicpera, Gro{\ss}, and
  G{\"u}nnemann(2020)}]{klicpera2020directional}
Klicpera, J.; Gro{\ss}, J.; and G{\"u}nnemann, S. 2020.
\newblock Directional message passing for molecular graphs.
\newblock \emph{arXiv preprint arXiv:2003.03123}.

\bibitem[{Landrum(2013)}]{landrum2013rdkit}
Landrum, G. 2013.
\newblock RDKit: A software suite for cheminformatics, computational chemistry,
  and predictive modeling.

\bibitem[{Li et~al.()Li, Sundar, Grigoryan, and Keating}]{literminator}
Li, A.~J.; Sundar, V.; Grigoryan, G.; and Keating, A.~E. ????
\newblock TERMinator: A Neural Framework for Structure-Based Protein Design
  using Tertiary Repeating Motifs.

\bibitem[{Li et~al.(2020)Li, Yan, Qiu, and Huang}]{li2020flat}
Li, X.; Yan, H.; Qiu, X.; and Huang, X. 2020.
\newblock FLAT: Chinese NER using flat-lattice transformer.
\newblock \emph{arXiv preprint arXiv:2004.11795}.

\bibitem[{Liu et~al.(2021)Liu, Wang, Liu, Zhang, Oztekin, and
  Ji}]{liu2021spherical}
Liu, Y.; Wang, L.; Liu, M.; Zhang, X.; Oztekin, B.; and Ji, S. 2021.
\newblock Spherical message passing for 3d graph networks.
\newblock \emph{arXiv preprint arXiv:2102.05013}.

\bibitem[{Lu et~al.(2019)Lu, Liu, Wang, Huang, Lin, and He}]{lu2019molecular}
Lu, C.; Liu, Q.; Wang, C.; Huang, Z.; Lin, P.; and He, L. 2019.
\newblock Molecular property prediction: A multilevel quantum interactions
  modeling perspective.
\newblock In \emph{Proceedings of the AAAI Conference on Artificial
  Intelligence}, volume~33, 1052--1060.

\bibitem[{Mackenzie, Zhou, and Grigoryan(2016)}]{mackenzie2016tertiary}
Mackenzie, C.~O.; Zhou, J.; and Grigoryan, G. 2016.
\newblock Tertiary alphabet for the observable protein structural universe.
\newblock \emph{Proceedings of the National Academy of Sciences}, 113(47):
  E7438--E7447.

\bibitem[{Martins et~al.(2012)Martins, Teixeira, Pinheiro, and
  Falcao}]{martins2012bayesian}
Martins, I.~F.; Teixeira, A.~L.; Pinheiro, L.; and Falcao, A.~O. 2012.
\newblock A Bayesian approach to in silico blood-brain barrier penetration
  modeling.
\newblock \emph{Journal of chemical information and modeling}, 52(6):
  1686--1697.

\bibitem[{Maziarka et~al.(2020)Maziarka, Danel, Mucha, Rataj, Tabor, and
  Jastrzebski}]{maziarka2020molecule}
Maziarka, {\L}.; Danel, T.; Mucha, S.; Rataj, K.; Tabor, J.; and Jastrzebski,
  S. 2020.
\newblock Molecule attention transformer.
\newblock \emph{arXiv preprint arXiv:2002.08264}.

\bibitem[{Maziarka et~al.(2021)Maziarka, Majchrowski, Danel, Gai{\'n}ski,
  Tabor, Podolak, Morkisz, and Jastrzebski}]{maziarka2021relative}
Maziarka, {\L}.; Majchrowski, D.; Danel, T.; Gai{\'n}ski, P.; Tabor, J.;
  Podolak, I.; Morkisz, P.; and Jastrzebski, S. 2021.
\newblock Relative Molecule Self-Attention Transformer.
\newblock \emph{arXiv preprint arXiv:2110.05841}.

\bibitem[{{\"O}zt{\"u}rk, {\"O}zg{\"u}r, and
  Ozkirimli(2018)}]{ozturk2018deepdta}
{\"O}zt{\"u}rk, H.; {\"O}zg{\"u}r, A.; and Ozkirimli, E. 2018.
\newblock DeepDTA: deep drug--target binding affinity prediction.
\newblock \emph{Bioinformatics}, 34(17): i821--i829.

\bibitem[{Ramakrishnan et~al.(2014)Ramakrishnan, Dral, Rupp, and von
  Lilienfeld}]{ramakrishnan2014quantum}
Ramakrishnan, R.; Dral, P.~O.; Rupp, M.; and von Lilienfeld, O.~A. 2014.
\newblock Quantum chemistry structures and properties of 134 kilo molecules.
\newblock \emph{Scientific Data}, 1.

\bibitem[{Ramakrishnan et~al.(2015)Ramakrishnan, Hartmann, Tapavicza, and
  Von~Lilienfeld}]{ramakrishnan2015electronic}
Ramakrishnan, R.; Hartmann, M.; Tapavicza, E.; and Von~Lilienfeld, O.~A. 2015.
\newblock Electronic spectra from TDDFT and machine learning in chemical space.
\newblock \emph{The Journal of chemical physics}, 143(8): 084111.

\bibitem[{Ramsundar et~al.(2015)Ramsundar, Kearnes, Riley, Webster, Konerding,
  and Pande}]{ramsundar2015massively}
Ramsundar, B.; Kearnes, S.; Riley, P.; Webster, D.; Konerding, D.; and Pande,
  V. 2015.
\newblock Massively multitask networks for drug discovery.
\newblock \emph{arXiv preprint arXiv:1502.02072}.

\bibitem[{Rong et~al.(2020)Rong, Bian, Xu, Xie, Wei, Huang, and
  Huang}]{rong2020self}
Rong, Y.; Bian, Y.; Xu, T.; Xie, W.; Wei, Y.; Huang, W.; and Huang, J. 2020.
\newblock Self-supervised graph transformer on large-scale molecular data.
\newblock \emph{arXiv preprint arXiv:2007.02835}.

\bibitem[{Sch{\"u}tt et~al.(2018)Sch{\"u}tt, Sauceda, Kindermans, Tkatchenko,
  and M{\"u}ller}]{schutt2018schnet}
Sch{\"u}tt, K.~T.; Sauceda, H.~E.; Kindermans, P.-J.; Tkatchenko, A.; and
  M{\"u}ller, K.-R. 2018.
\newblock Schnet--a deep learning architecture for molecules and materials.
\newblock \emph{The Journal of Chemical Physics}, 148(24): 241722.

\bibitem[{Sch{\"u}tt, Unke, and Gastegger(2021)}]{schutt2021equivariant}
Sch{\"u}tt, K.~T.; Unke, O.~T.; and Gastegger, M. 2021.
\newblock Equivariant message passing for the prediction of tensorial
  properties and molecular spectra.
\newblock \emph{arXiv preprint arXiv:2102.03150}.

\bibitem[{Somnath, Bunne, and Krause(2021)}]{somnath2021multi}
Somnath, V.~R.; Bunne, C.; and Krause, A. 2021.
\newblock Multi-Scale Representation Learning on Proteins.
\newblock In \emph{Thirty-Fifth Conference on Neural Information Processing
  Systems}.

\bibitem[{Subramanian et~al.(2016)Subramanian, Ramsundar, Pande, and
  Denny}]{subramanian2016computational}
Subramanian, G.; Ramsundar, B.; Pande, V.; and Denny, R.~A. 2016.
\newblock Computational modeling of $\beta$-secretase 1 (BACE-1) inhibitors
  using ligand based approaches.
\newblock \emph{Journal of chemical information and modeling}, 56(10):
  1936--1949.

\bibitem[{Thomas et~al.(2018)Thomas, Smidt, Kearnes, Yang, Li, Kohlhoff, and
  Riley}]{thomas2018tensor}
Thomas, N.; Smidt, T.; Kearnes, S.; Yang, L.; Li, L.; Kohlhoff, K.; and Riley,
  P. 2018.
\newblock Tensor field networks: Rotation-and translation-equivariant neural
  networks for 3d point clouds.
\newblock \emph{arXiv preprint arXiv:1802.08219}.

\bibitem[{Townshend et~al.(2020)Townshend, V{\"o}gele, Suriana, Derry, Powers,
  Laloudakis, Balachandar, Anderson, Eismann, Kondor
  et~al.}]{townshend2020atom3d}
Townshend, R.~J.; V{\"o}gele, M.; Suriana, P.; Derry, A.; Powers, A.;
  Laloudakis, Y.; Balachandar, S.; Anderson, B.; Eismann, S.; Kondor, R.;
  et~al. 2020.
\newblock ATOM3D: Tasks On Molecules in Three Dimensions.
\newblock \emph{arXiv preprint arXiv:2012.04035}.

\bibitem[{Vaswani et~al.(2017)Vaswani, Shazeer, Parmar, Uszkoreit, Jones,
  Gomez, Kaiser, and Polosukhin}]{vaswani2017attention}
Vaswani, A.; Shazeer, N.; Parmar, N.; Uszkoreit, J.; Jones, L.; Gomez, A.~N.;
  Kaiser, {\L}.; and Polosukhin, I. 2017.
\newblock Attention is all you need.
\newblock In \emph{Advances in neural information processing systems},
  5998--6008.

\bibitem[{Vinyals, Bengio, and Kudlur(2015)}]{vinyals2015order}
Vinyals, O.; Bengio, S.; and Kudlur, M. 2015.
\newblock Order matters: Sequence to sequence for sets.
\newblock \emph{arXiv preprint arXiv:1511.06391}.

\bibitem[{Wang et~al.(2005)Wang, Fang, Lu, Yang, and Wang}]{wang2005pdbbind}
Wang, R.; Fang, X.; Lu, Y.; Yang, C.-Y.; and Wang, S. 2005.
\newblock The PDBbind database: methodologies and updates.
\newblock \emph{Journal of medicinal chemistry}, 48(12): 4111--4119.

\bibitem[{Wang et~al.(2018)Wang, Girshick, Gupta, and He}]{wang2018non}
Wang, X.; Girshick, R.; Gupta, A.; and He, K. 2018.
\newblock Non-local neural networks.
\newblock In \emph{Proceedings of the IEEE conference on computer vision and
  pattern recognition}, 7794--7803.

\bibitem[{Wernicke(2006)}]{wernicke2006efficient}
Wernicke, S. 2006.
\newblock Efficient detection of network motifs.
\newblock \emph{IEEE/ACM transactions on computational biology and
  bioinformatics}, 3(4): 347--359.

\bibitem[{Wu et~al.(2020)Wu, Liu, Lin, Lin, and Han}]{wu2020lite}
Wu, Z.; Liu, Z.; Lin, J.; Lin, Y.; and Han, S. 2020.
\newblock Lite transformer with long-short range attention.
\newblock \emph{arXiv preprint arXiv:2004.11886}.

\bibitem[{Wu et~al.(2018)Wu, Ramsundar, Feinberg, Gomes, Geniesse, Pappu,
  Leswing, and Pande}]{wu2018moleculenet}
Wu, Z.; Ramsundar, B.; Feinberg, E.~N.; Gomes, J.; Geniesse, C.; Pappu, A.~S.;
  Leswing, K.; and Pande, V. 2018.
\newblock MoleculeNet: a benchmark for molecular machine learning.
\newblock \emph{Chemical science}, 9(2): 513--530.

\bibitem[{Xiong et~al.(2019)Xiong, Wang, Liu, Zhong, Wan, Li, Li, Luo, Chen,
  Jiang et~al.}]{xiong2019pushing}
Xiong, Z.; Wang, D.; Liu, X.; Zhong, F.; Wan, X.; Li, X.; Li, Z.; Luo, X.;
  Chen, K.; Jiang, H.; et~al. 2019.
\newblock Pushing the boundaries of molecular representation for drug discovery
  with the graph attention mechanism.
\newblock \emph{Journal of medicinal chemistry}, 63(16): 8749--8760.

\bibitem[{Xu et~al.(2017)Xu, Wang, Zhu, and Huang}]{xu2017seq2seq}
Xu, Z.; Wang, S.; Zhu, F.; and Huang, J. 2017.
\newblock Seq2seq fingerprint: An unsupervised deep molecular embedding for
  drug discovery.
\newblock In \emph{Proceedings of the 8th ACM international conference on
  bioinformatics, computational biology, and health informatics}, 285--294.

\bibitem[{Yang et~al.(2019)Yang, Swanson, Jin, Coley, Eiden, Gao, Guzman-Perez,
  Hopper, Kelley, Mathea et~al.}]{yang2019analyzing}
Yang, K.; Swanson, K.; Jin, W.; Coley, C.; Eiden, P.; Gao, H.; Guzman-Perez,
  A.; Hopper, T.; Kelley, B.; Mathea, M.; et~al. 2019.
\newblock Analyzing learned molecular representations for property prediction.
\newblock \emph{Journal of chemical information and modeling}, 59(8):
  3370--3388.

\bibitem[{Ying et~al.(2021)Ying, Cai, Luo, Zheng, Ke, He, Shen, and
  Liu}]{ying2021transformers}
Ying, C.; Cai, T.; Luo, S.; Zheng, S.; Ke, G.; He, D.; Shen, Y.; and Liu, T.-Y.
  2021.
\newblock Do Transformers Really Perform Bad for Graph Representation?
\newblock \emph{arXiv preprint arXiv:2106.05234}.

\bibitem[{Zhang et~al.(2020)Zhang, Hu, Subramonian, and Sun}]{zhang2020motif}
Zhang, S.; Hu, Z.; Subramonian, A.; and Sun, Y. 2020.
\newblock Motif-driven contrastive learning of graph representations.
\newblock \emph{arXiv preprint arXiv:2012.12533}.

\bibitem[{Zhang et~al.(2021)Zhang, Liu, Wang, Lu, and Lee}]{zhang2021motif}
Zhang, Z.; Liu, Q.; Wang, H.; Lu, C.; and Lee, C.-K. 2021.
\newblock Motif-based Graph Self-Supervised Learning for Molecular Property
  Prediction.
\newblock \emph{arXiv preprint arXiv:2110.00987}.

\end{thebibliography}
